\documentclass[twocolumn,prl, superscriptaddress,floatfix, showpacs]{revtex4}

\usepackage{graphicx}
\usepackage{amsmath} 
\usepackage{dcolumn} 
\usepackage{amsfonts}
\usepackage{amssymb}
\usepackage{amscd}

\usepackage{color}

\newcommand{\red}{\color{black}}
\newcommand{\Red}{\color{black}}  
\definecolor{brown}{rgb}{0.8,0.2,0.1}
\newcommand{\brown}{\color{brown}}

\newcommand{\su}[1]{\mathfrak{su}(#1)}

\begin{document}



\title{ Eigenstate Estimation for the Bardeen-Cooper-Schrieffer (BCS) Hamiltonian}


\author{S.Y.\ Ho}
\affiliation{Department of Physics, University of Toronto, Toronto, 
Ontario M5S 1A7, Canada}
\affiliation{Centre for Quantum Technologies and Department of Physics,
National University of Singapore, 3 Science Drive 2, 117543 Singapore}
\affiliation{Department of Physics, University of Illinois at Urbana-Champaign,
1110 West Green Street, Urbana, Illinois 61801-3080, USA}
\author{D.J.~Rowe}
\affiliation{Department of Physics, University of Toronto, Toronto,
Ontario M5S 1A7, Canada}
\author{S. De Baerdemacker}
\affiliation{Department of Physics, University of Toronto, Toronto,
Ontario M5S 1A7, Canada} 
\affiliation{{Ghent University, Department of Physics and Astronomy,
Proeftuinstraat 86, B-9000 Gent, Belgium}}
\affiliation{Department of Physics, University of Notre Dame, Notre Dame,
Indiana 46556-5670, USA}

\date{\today}

\begin{abstract}
We show how multi-level BCS Hamiltonians of finite systems in the 
strong pairing interaction regime can be accurately approximated using 
multi-dimensional shifted harmonic oscillator Hamiltonians. 
In the Shifted Harmonic Approximation (SHA), discrete quantum 
state variables are approximated as continuous ones and 
algebraic Hamiltonians are replaced by differential operators.
Using the SHA, the results of the BCS theory, such as the gap equations, 
can be easily derived without the BCS approximation.
In addition, the SHA preserves the symmetries 
associated with the BCS Hamiltonians. Lastly, for all interaction strengths, 
the SHA can be used to identify the most important basis states -- allowing
accurate computation of low-lying eigenstates by 
diagonalizing BCS Hamiltonians in small subspaces of what may otherwise 
be vastly larger Hilbert spaces.

\end{abstract}



\maketitle


The traditional method of finding eigenvalues of a Hamiltonian 
$\hat H (\{\hat X_\nu\})$ (expressed as a polynomial in the elements  
$\{\hat X_\nu\}$ of a Lie algebra $\mathfrak{g}$) is by diagonalization.  
However, in realistic many-body systems the Hamiltonian matrices can be huge.
The problem is then to find an approximation such that the salient features 
of the Hamiltonian are retained.
In this letter, the so-called Shifted Harmonic Approximation (SHA), 
introduced by Chen et al.\ \cite{SHA}, is developed and extended to many 
degrees of freedom.
The key principle behind the SHA is to replace
discrete quantum state variables by continuous ones. 
Algebraic Hamiltonians are then replaced by differential operators.
This approach offers new insights even for well studied
systems such as those with a Bardeen-Cooper-Schrieffer (BCS) 
Hamiltonian \cite{BCS, Kerman}, which in general cannot be
solved exactly. The traditional BCS approximation provides accurate 
results in the thermodynamic limit but violates
particle-number conservation. For finite systems, this
is a major source of inaccuracies but its effects can be reduced by 
number conserving extensions of the BCS theory
such as \cite{NCBCS1, NCBCS2}.

Recent studies of superconductivity in metallic nano-grains \cite{NG}
and atomic nuclei \cite{RGnucleus} have led to a revival of interest 
in the Richardson-Gaudin approach \cite{richardson, gaudin:76}.  
Classes of BCS Hamiltonians with level-independent interactions
are shown to be integrable and solvable by means of an algebraic Bethe ansatz. 
However, the numerical solutions are challenging to compute \cite{RG2} 
and the eigenstates are not easy to use. 
Moreover, among the set of BCS Hamiltonians, 
there is only a small number of special cases \cite{dukelsky:01}
that are solvable by the Richardson-Gaudin method.

Here, using the SHA which is number conserving, we show that a general 
$k$-level BCS Hamiltonian can be approximated as a $(k-1)$-dimensional 
shifted oscillator Hamiltonian. Accurate approximations of the low-lying 
eigenstates are then easily obtained in the strong interaction regime.
In the weak interaction regime, the SHA can also be used to identify the most 
important basis states for computing the low-lying eigenstates accurately.

Consider an irreducible representation (irrep) of the $\su{2}$ algebra 
on the Hilbert space spanned by basis states 
$\{|m\rangle, m = -j, -j+1 \dots, j \}$. 
Any state $|\phi\rangle$ in this Hilbert space,
e.g., an eigenstate of a Hamiltonian in the $\su{2}$ algebra,
can be expressed as a linear combination of the basis states
$|\phi\rangle =\sum_m |m\rangle\langle m|\phi\rangle = 
\sum_m |m\rangle \phi(m)$, where the coefficient 
$\phi(m) = \langle m|\phi\rangle$
is a discrete distribution of $m$.
The action of the $\su{2}$ operators on such a distribution,
defined by 
$ \hat{\mathcal{J}}_k \phi(m) =  \langle m | \hat{J}_k |\phi\rangle$, 
is then
\begin{align} 
\hat{\mathcal{J}}_z \phi(m) &=   
m\phi(m)\label{continuation:Jz} ,\\ 
\hat{\mathcal{J}}_\pm \phi(m)
&=  \sqrt{(j \mp m + 1)(j \pm m)} \, \phi(m\mp 1). 
\label{continuation:Jpm}
\end{align}
For large values of $j$ and for a state for which $\phi(m)$ varies slowly
with the discrete variable $m$, we can now make the continuous 
variable approximation of extending $m$ to continuous values and replacing
$\phi(m)$ by a smooth function
{\Red $\psi(x)$, defined such that $\psi(x)=\phi(m)$ when $x=m/j$.} 
We can then use the identity
$\psi(x\mp\frac1j)=\exp{(\mp\frac{1}{j} \frac{d}{dx})}\psi(x)$ 
and, assuming the expansion of $\exp{(\mp\frac{1}{j} \frac{d}{dx})}\psi(x)$ 
to be rapidly convergent, make the approximation
\begin{eqnarray} 
\hat{\mathcal{J}}_\pm \psi(x)
&=& j\sqrt{(1 \mp x + \tfrac1j)(1 \pm x)}\,
\exp\!\Big(\mp \frac{1}{j}\frac{d}{dx}\Big)\psi(x) \nonumber\\
&\approx&   j\sqrt{1-x^2} \, 
\left[1 \mp \frac{1}{j} \frac{d}{dx}  + \frac{1}{2j{^2}} \frac{d^2}{dx^2} 
\right]  \psi(x).
\label{Jpmdiff}
\end{eqnarray}
Note that we have omitted the $1/j$ term to obtain 
$\sqrt{1-x^2}$ in eq.~(\ref{Jpmdiff}). 
This term is negligible for large values of $j$,
but could be included in a more complete calculation.
If the function $\psi(x)$ is  
(i) slow varying, 
(ii) localized about a value $x_o$ and 
(iii) vanishes when $|x|\rightarrow 1$, we can make the 
\emph{Shifted Harmonic Approximation} (SHA). 
In this approximation, the action of an $\su{2}$ operator, such as 
$\hat{\mathcal{J}}_\pm$ in eq.\,(\ref{Jpmdiff}), on $\psi(x)$
is obtained by expanding it about $x_o$ up to bilinear terms.  
Similarly, a Hamiltonian that is quadratic in the elements of an 
$\su{2}$ algebra and has low-lying eigenfunctions that satisfy the SHA criteria
can be mapped to a harmonic oscillator Hamiltonian  
$\hat{\mathcal{H}}_\textrm{SHA}$ that is bilinear in $(x-x_o)$ and $d/dx$. 

Now consider a \emph{multi-level} BCS Hamiltonian consisting of
fermions in $k$ single-particle energy levels. 
The operators $a^\dag_{\mu_p}$ ($a_{\mu_p}$)
create (annihilate) a fermion in a state $\mu_p$ at level $p$, and the 
operators for the corresponding time-reversed states are 
$a^\dag_{\bar{\mu}_p}$ ($a_{\bar{\mu}_p}$).  As shown 
by Kerman \emph{et.\ al.}\ \cite{Kerman}, these operators can be 
combined to form $\su{2}$ \emph{quasi-spin operators}
\begin{equation}
\hat{J}^p_z = \tfrac12\!\sum_{\mu_p>0} ( a^\dag_{\mu_p} a_{\mu_p} - 
a_{\bar{\mu}_p} a^\dag_{\bar{\mu}_p}), \;\; 
\hat{J}_+^p=\tfrac{1}{2}\!\sum_{\mu_p} a^\dag_{\mu_p}a^\dag_{\bar{\mu}_p}. 
\end{equation}
Here, the operator $\hat{J}^p_+$ creates a pair of particles in 
time-reversed states at level $p$.
Together with the pair annihilation operators,  $J_-^p=(J_+^p)^\dag$,
these quasi-spin operators belong to an $\su{2}\oplus \su{2} \oplus \ldots$ 
algebra with commutation relations
$[\hat{J}^p_+ , \hat{J}^q_- ]=2\hat{J}^p_z\delta_{pq}$ and
$[\hat{J}^p_z , \hat{J}^q_\pm]=\pm\hat{J}^p_\pm\delta_{pq}.$
In this formalism, the BCS Hamiltonian is written as
\begin{equation}\label{PH}
\hat{H} = \sum_{p=1}^k \epsilon_p \hat{n}^p - 
\sum_{p,q}^k G_{pq} \hat{J}^p_ + \hat{J}^q_-,
\end{equation}
where $\hat{n}^p=2(\hat{J}^p_z+ j_p)$ is the particle number operator for 
the level with single particle energy $\epsilon_p$. The operator 
$\hat{J}^p_+\hat{J}^q_-$ scatters a pair of particles from level $q$ to 
level $p$ and $G_{pq}$ is the corresponding interaction strength.  

The Hamiltonian (\ref{PH}) conserves both particle number and the 
number of paired particles.  
Without loss of generality, we consider a system with no 
unpaired particles. For level $p$, let $\{ |j_pm_p\rangle\}$ 
denote the basis states for the irreducible $\su{2}_p$ representation
for which $|j_p,m_p\!=\! -j_p\rangle$ is the zero-pair state and 
$m_p$ increases by one with every added pair to 
reach the value $m_p=j_p$, when the level is completely filled.  
Basis states for the  $k$-level pairing model with no unpaired particles, 
are then defined by  
$|\mathbf{m}\rangle = |j_1m_1\rangle \otimes |j_2m_2\rangle \otimes
\dots \otimes  |j_km_k\rangle$.

\begin{figure}[th]
	\centering \vspace{-0. in}
		\centerline{\includegraphics[width=3.2 in]{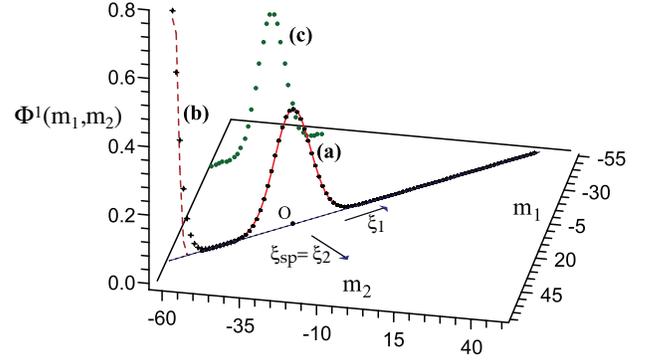}} 
	\caption{(Color online) A sample two-level system. 
The components of the exact ground state eigenvectors 
($\Phi^1(m_1,m_2)$) are indicated with dots on an 
\Red $(m_1,m_2)$ plane for (a) large and (b) small interactions.  
The solid and dash lines are the corresponding SHA eigenfunctions.
The ground state for a different $N$ is shown by (c).
The directions of the transformed coordinates, $\xi_1$ and $\xi_{sp} = \xi_2$,
with their origins at the point O $\sim (j_1x_{o1}, j_2x_{o2})$
are indicated for (a).}
	\label{2lvl}
\end{figure}

Eigenstates $|\Phi^i \rangle$ of the pairing Hamiltonian (\ref{PH}), 
with a fixed pair number $N$, are given by linear combinations of 
the basis states,  $|\mathbf{m}\rangle$, for which 
$\sum_{p=1}^km_p=N-\tfrac{1}{2}N_\textrm{max}$ where $N_\textrm{max}$ 
is the maximum number of pairs possible in the system.  If we plot the set of 
allowed basis states $|\mathbf{m}\rangle$ for the $N$-pair system as points on
an $\{m_1,m_2,\dots,m_k\}$ grid, these points lie on a $(k-1)$-dimensional 
hyperplane. The direction orthogonal to this hyperplane is described as 
\emph{spurious} because there is no dynamics associated with it when $N$ 
is fixed.  See FIG.~\ref{2lvl} for a sample $2$-level system. 

To apply the SHA to the pairing Hamiltonian (\ref{PH}), 
we define $\hat{\mathcal{H}}\Phi^i(\mathbf{m}) 
=\langle {\mathbf m}|\hat{H} | \Phi^i\rangle$.
Then, assuming that the low-lying eigenfunctions $\Phi^i(\mathbf{m})$ vary
slowly with $m$, we define $k$ continuous variables $x_p=m_p/j_p$, and the 
continuous eigenfunction $\Psi^i(\mathbf{x})$ as before.
Assuming also that the wave functions, $\Psi^i(\mathbf{x})$, 
are localized around a point $\mathbf{x}_o$, 
and that the criteria for the validity of the SHA, 
listed above, are satisfied, 
we expand  the Hamiltonian $\hat{\mathcal{H}}$ up to 
bilinear terms in  $x'_p = (x_p -  x_{op})$ and 
$ \frac{\partial}{\partial x'_p}$ {\red to} obtain
\begin{align} \label{SHAPH}
\hat{\mathcal{H}} \, \approx \,\, & \hat{\mathcal{H}}_\textrm{SHA} =  
\sum_{p,q}^k\left[- {{\frac1{j_p}}\frac{\partial}{\partial x^\prime_p} 
A_{pq}{{\frac1{j_q}}}\frac{\partial}{\partial x^\prime_q}}
+{j_p} x'_p B_{pq} j_qx'_q\right]\nonumber\\
&\qquad \qquad +\sum_p^k D_p j_px'_p + {E},
\end{align}
which is essentially the Hamiltonian for a coupled $k$-dimensional harmonic 
oscillator with the origin shifted to $x_{op}$. 
It contains: 
an inverse mass tensor $A_{pq}$,  a spring constant tensor $B_{pq}$, a set 
of shifts $D_p$  and  a constant $E$. Their components, defined in terms of 
$\kappa_r = (1-x_{or}^2)$ and $T_{pq} = G_{pq}j_pj_q\sqrt{\kappa_p\kappa_q}$, 
are given by
\begin{align}
&A_{pq}= \sum_r T_{pr}\delta_{pq} - T_{pq} ,\label{Apq}\\
&B_{pq}= \sum_r T_{pr}\frac{\delta_{pq}}{{ j_p^2}\kappa_p^2} - 
T_{pq}\frac{x_{op}x_{oq}} {{ j_pj_q}\kappa_p\kappa_q},\label{Bpq}\\
&D_p = 2\epsilon_p  - {G_{pp}} + \frac{2x_{op}}{j_p\kappa_{p}} 
\sum_{r} T_{pr},\label{Dq} \\
&E_{}= \sum_p  j_p(2\epsilon_p - G_{pp})(1+x_{op}) - \sum_{p,q} T_{pq}
\label{Eq}.
\end{align}
The conservation of particle number in the SHA formalism is verified by showing 
that the SHA representation of the number operator
$\hat n = 2\sum_p (\hat J^p_z + j_p)$ commutes with 
$\hat{\mathcal{H}}_\textrm{SHA}$.  Thus, it is appropriate to 
make a change of variables, $j_px^\prime_p \to  \xi_i$, such that $\xi_k$ 
(denoted $\xi_{sp}$ in FIG.~\ref{2lvl}) is an $N$-dependent constant 
and the Hamiltonian (\ref{SHAPH}) becomes that of a system of $(k\!-\!1)$
harmonic oscillators,
\begin{equation} \label{SHAfinal}
\hat{\mathcal{H}}_\textrm{SHA} = 
\sum_{i=1}^{k-1}\left(-\alpha_{i}\frac{\partial^2}{\partial\xi_{i}^{2}}
+\beta_{i}\xi_{i}^{2}\right) + E.
\end{equation}

Observe that while the dynamics of the oscillator is $(k-1)$-dimensional, 
the tensors $A$ and $B$ are $k$-dimensional. However, 
the inverse mass tensor $A$ is determined to have an eigenvector parallel 
to the spurious direction with zero eigenvalue; 
this implies that the corresponding vibrational mass is infinite, 
consistent with $N$ being a good quantum number in the SHA. 
To evaluate the quantities in eqs.\ (\ref{Apq} - \ref{Eq}), we need  
the numerical values of $x_{op}$.
The point ${\bf x_o}$ is naturally defined as the minimum of the harmonic 
oscillator potential on the $(k-1)$-dimensional hyperplane 
(see FIG.~\ref{2lvl}). 
Once the pair number $N$ is selected,  the numerical values of 
$x_{op}$ are determined by the $D_p$ shift functions  of Eq. (\ref{Dq}).

Having determined $x_{op}$, we can use the transformation that 
diagonalizes $A$ to obtain the dynamics on the hyperplane.
The properties of the $(k-1)$-dimensional oscillator on the hyperplane
are solved using normal-mode theory.
The eigenvalues of the SHA Hamiltonian (\ref{SHAfinal}) are
\begin{equation} \label{SHAEa}
\mathcal{E}_\nu = \sum_{i=1}^{k-1} (\nu_i + \tfrac12)\omega_i + E,
\end{equation} 
where $\omega_i = 2\sqrt{{\alpha_i}{\beta_i}}$ and $\nu=\{\nu_i\}$ is a set 
of integers indicating the number of oscillator quanta in mode $i$.
The corresponding set of {\em SHA eigenfunctions} are
\begin{equation} \label{SHAwf}
\Psi_\nu ({{\bf \xi}}) = \eta\prod_{i=1}^{k-1} \big(2^{\nu_i} {\sigma_i} 
\nu_i!\sqrt{\pi}\big)^{-\frac12}
H_{\nu_i}({\tfrac{{\xi_i}}{\sigma_i})} e^{-\frac12 (\frac{{\xi_i}}{\sigma_i})^2} 
\end{equation}
where $H_{\nu_i}$ are Hermite polynomials, 
$\sigma_i = \sqrt[4]{\frac{\alpha_i}{\beta_i}}$ 
are the SHA widths, and $\eta$ is a normalization factor.  
From here, we can approximate the 
coefficients $\Phi^i(\mathbf{m})=\langle\mathbf{m}|\Phi^i\rangle$, and hence 
the eigenfunction, in the original discrete basis by 
evaluating $\Psi_\nu$ at the points ${\bf \xi}$ corresponding to $\mathbf{m}$.
We refer to these approximate eigenfunctions of the Hamiltonian 
$\hat H$, as the {\em SHA basis}.  

It is worth noting that the quantity $\tfrac12(x_{op}+ 1)$ in the SHA 
can be interpreted as the mean fractional occupancy of a single-particle 
level $p$ in parallel with $v_p^2$ in BCS theory \cite{BCS}.  
Similarly, the SHA shift equations for $x_{op}$ correspond to the 
BCS gap equations for $v_p^2$.  In addition, the SHA energy $E$ is almost 
identical to the BCS ground state energy.  
It has been shown, in several model calculations, that including higher order 
corrections in $\frac{1}{j}$ lowers the SHA ground-state energy in 
the strong interaction regime below that of the BCS approximation.
Thus, we   obtain an insightful interpretation of the SHA treatment of the 
pairing model as an extension of the BCS method to a number conserving 
approximation which takes account of the fluctuations of the particle number 
in each single-particle level about its mean BCS value.

The transformed $\xi$-coordinates for a 2-level model are illustrated in 
{FIG.~\ref{2lvl}}. The SHA eigenfunction (line) corresponding to a large 
interaction (compared to single particle energies spacing), indicated as 
(a), are in good agreement with the exact components of the eigenvectors 
given by diagonalization. Similar accuracy is obtained for the next few 
higher-energy states (not shown). 
If greater precision is required, an even more accurate description 
of the low-lying eigenstates can be obtained by diagonalizing the BCS 
Hamiltonian in a subspace spanned by a small number of SHA basis states.
For  weaker interactions, as in (b), the SHA does not predict the components of 
the eigenvectors accurately as in (a).
This is the regime in which the conditions for the validity of the SHA 
are not well satisfied. Nevertheless, some
SHA predictions, such as $x_{op}$ and $\sigma_i$, remain accurate 
- a subtlety not yet fully understood. Thus, we can use
these predictions to identify a small subset of basis states that contribute 
significantly to the low-lying eigenstates in the weak interaction regime
and also obtain  very accurate results for them
by diagonalizing small Hamiltonian matrices.
 
To illustrate the effectiveness of the SHA, 
we consider a system with four degenerate single-particle energy levels. 
This  relatively small system is selected so that exact eigenstates can 
be obtained by diagonalization.  Based on other applications of the SHA
\cite{RdeG}, we expect the SHA to be even more accurate and effective 
in application to systems of single-particle levels of higher multiplicities.

\begin{figure}[ht!]
	\centering
		\includegraphics[scale=0.415]{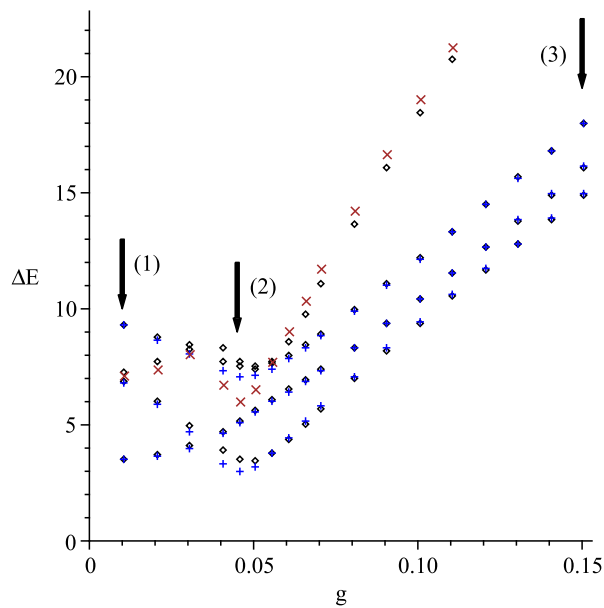}
	\caption{(Color online) The SHA predicted excitation energies for
	 $\{n_1, n_2, n_3\}$=$\{1,0,0\}, \{0,1,0\}$ and $\{0,0,1\}$ are indicated 
	 with `+' and those for $\{2,0,0\}$ are indicated with `{\brown $\times$}'.	
	 The exact lowest four excitation energies are given by `$\diamond$'.
	 We mark three points of interest (1) $g=0.010$, (2) $g=0.045$ and 
	 (3) $g=0.150$. Note that the SHA excitation energies corresponding 
	 to $\{2,0,0\}$ cross-over those of 	$\{0,0,1\}$ at $g \sim 0.05$, 
	 indicating the system transiting from one regime to another.}
	\label{fig:PHOm}
\end{figure}

Exact and SHA-estimated excitation energies of a sample 
$4$-level model with $N=28$ pairs,  $j=[7,8,9,10]$ and 
$\epsilon=[0.5,2.3,6.1,7.3]$ are shown in FIG.~\ref{fig:PHOm}. 
A simple arbitrary rule for $G$ is used: 
$G_{pq}= (2.0 - 0.1|\epsilon_p - \epsilon_q|)g$, where $g$ controls the 
interaction strength.  Exact results are obtained by diagonalizing the 
$3231\times 3231$ Hamiltonian matrix.
For this 4-level model, the SHA oscillator in the strong interaction regime 
is 3-dimensional.  The number of oscillator quanta in each mode 
is given by $\{n_1, n_2, n_3\}$. The excitation energy 
$\Delta \mathcal{E}_{ \{n_1, n_2, n_3\}} = \mathcal{E}_{ \{n_1, n_2, n_3\}} 
- \mathcal{E}_{\{0, 0, 0\}}$ for the low-lying states are shown.
From the figure, we see that the SHA-predicted excitation energies 
(`+', `{\brown $\times$}') are in good agreement with the exact results 
from diagonalization (`$\diamond$') for a wide range of interactions.
Note that while the SHA excitation energies for $g \sim 0.05$ are less accurate 
than for other values of $g$, the trends in how
the excitation energies vary with interaction strength
are still closely captured by the SHA.

\begin{table}[th]

	\caption{The lowest few excitation energies for (1) $g=0.010$ (2) $g=0.045$
	  and (2) $g=0.150$ as predicted by the SHA and diagonalizing in 
	  the most relevant subspaces (indicated by $Diag$) are shown.  
	  The `$Diag$' results are given up to the digit that agrees with the 
	  corresponding exact result. All SHA predictions are given to two decimal 
	  places in (\,\,) with the number of oscillator quanta given in \{\, \,\} in 
	  the subscript.}	
	\label{tab:Results}
	
	\centering \setlength{\extrarowheight}{2 pt}
		\begin{tabular}{ccc}\hline
		           (1) $g=0.010$  		      & (2) $g=0.045$ 	& (3) $g=0.150$ \\
$Diag_{\su{2}}$ (SHA)    & $Diag_{\su{2}}$ (SHA)   & $Diag_\textrm{SHA}$(SHA) \\ 
\hline
3.650313 (3.61)$_{ \{1,0,0\}}$				& 3.60 (3.07)$_{ \{1,0,0\}}$     
& 15.03 (15.06)$_{ \{1,0,0\}}$ \\     
6.9484 (6.90)$_{ \{0,1,0\}}$					& 5.258 (5.19)$_{ \{0,1,0\}}$    
& 16.18 (16.22)$_{ \{0,1,0\}}$ \\
7.38050 (7.21)$_{ \{2,0,0\}}$	      & 7.65 (6.14)$_{ \{2,0,0\}}$     
& 18.1 (18.08)$_{ \{0,0,1\}}$ \\
9.4210 (9.38)$_{ \{0,0,1\}}$					& 7.8 (7.20)$_{ \{0,0,1\}}$   
& 29.5  (30.12)$_{ \{2,0,0\}}$ \\
10.5531  (10.51)$_{ \{1,1,0\}}$ &   8.37	 (8.26)$_{ \{1,1,0\}}$    
&   30.7 (31.28)$_{ \{1,1,0\}}$\\
\hline
\end{tabular}
\end{table} 
Lastly, we show in TABLE \ref{tab:Results} the low-lying excitation energies 
obtained  from the SHA and by diagonalizing the BCS Hamiltonian 
in the space spanned by the most important basis states identified by the SHA.
For points (1) and (2) in FIG.~\ref{fig:PHOm}, the lowest $50$ and $300$ 
$\su{2}$ basis states are used respectively. 
For point (3), we used the lowest 286 SHA wave functions as a basis.

The results obtained, cf.\ last column of TABLE \ref{tab:Results}, 
show the SHA to be very successful for deriving low-energy spectra 
of BCS Hamiltonians in the strong interaction regime 
in which it is most valid.  
They also show the SHA  to be a good first-order approximation in general.
As TABLE \ref{tab:Results} indicates, accurate results can be obtained 
for any interaction by using the SHA to select relatively small subsets 
of basis states for diagonalizations. 

The SHA predicts essentially the same mean level occupancies in the 
ground state as the BCS approximation. In addition, it gives the fluctuations 
in these occupancies in a manner that conserves particle number. It also 
conserves the $\su{2} \otimes \su{2} \ldots$ symmetry of the 
pairing Hamiltonian  (defined by the values of the ${j_p}$ quantum numbers).
The SHA gives the low-energy states of all irreps of this symmetry group.  
This is in contrast to the BCS approximation which is only designed 
to give an approximation for the ground state and  
quasi-particle approximations for the low-energy states of 
neighbouring odd-particle systems. 
States of maximal $\su{2} \otimes \su{2} \ldots$ symmetry are 
\emph{unbroken-pair} states, whereas the \emph{broken-pair} states of other 
irreps have unpaired particles in one or more single-particle levels.  
This reduces the number of states available to the paired particles so that 
these irreps are obtained by reducing the quasi-spin of each level $p$ 
by the replacement $j_p \to j_p-\tfrac12$ for each unpaired particle 
in the level.  The states of such irreps are handled in the same way 
in the SHA, except for the different values of the quasi-spins.
To conclude, we note the significant possibility that the 
continuous variable approximation, underlying the SHA, has the potential 
to be applied to derive other solvable differential equations. 
This potential remains to be explored.

The authors wish to thank Veerle Hellemans and Trevor Welsh for their 
discussions. SDB is an ``FWO-Vlaanderen'' post-doctoral researcher and 
acknowledges an FWO travel grant for a ``long stay abroad'' at the 
University of Toronto and the University of Notre Dame.  SYH acknowledges 
the funding from the National Research Foundation and the Ministry of 
Education (Singapore).  This work was supported in part by the Natural 
Sciences and Engineering Research Council of Canada.

\end{document}